\documentclass[a4paper,11pt]{article}
\pdfoutput=1
\usepackage[utf8]{inputenc}
\usepackage[english]{babel}
\usepackage[T1]{fontenc}
\usepackage{authblk}
\usepackage{amsmath}
\usepackage{amssymb}
\usepackage{hyperref}
\usepackage{orcidlink}
\usepackage{caption}
\usepackage{graphicx}
\usepackage{booktabs}
\usepackage{tabularx}
\usepackage{array}

\graphicspath{{figures/}{./}}

\usepackage{amsmath,amssymb,amsfonts}
\usepackage[ruled]{algorithm}
\usepackage{algpseudocode}

\newcommand{\trms}{t_{\mathrm{rms}}}

\newcommand{\Id}{\mathbb{I}}

\usepackage{tikz}
\usepackage{lipsum}

\newcommand{\re}{\textrm{Re}}

\newcommand{\ket}[1]{|#1\rangle}

\begin{document}

\title{Rodeo Filtering for Direct Steady-State Estimation in Open Quantum Systems}

\renewcommand{\Authfont}{\normalsize}
\renewcommand{\Affilfont}{\small}
\setlength{\affilsep}{0.4em}

\author[1,2]{Hyeonjun Yeo~\orcidlink{0009-0005-9368-306X}$^*$}
\author[2,3]{Jongin Jeong~\orcidlink{0009-0009-4658-2724}$^*$}
\author[1]{Soyoung Shin~\orcidlink{0009-0005-7449-196X}}
\author[4,5]{Ha Eum Kim~\orcidlink{0009-0004-2614-0834}$^\dagger$}

\affil[1]{Department of Physics and Astronomy, Seoul National University, Seoul 08826, Korea}
\affil[2]{Team QST, Seoul National University, Seoul 08826, Korea}
\affil[3]{Department of Physics, Pusan National University, Busan, 46241, Korea}
\affil[4]{Department of Electrical and Computer Engineering, University of Illinois at Urbana-Champaign, Urbana, IL 61801, USA}
\affil[5]{Department of Mathematics and Research Institute for Basic Sciences, Kyung Hee University, Seoul, 02447, Korea}

\affil[ ]{\small $^*$ These authors contributed equally to this work.}
\affil[ ]{\small $^\dagger$ Corresponding author.}

\affil[ ]{\small
H.Y., \href{mailto:duguswns11@snu.ac.kr}{duguswns11@snu.ac.kr};
J.J., \href{mailto:jijeong2004@gmail.com}{jijeong2004@gmail.com};
H.E.K., \href{mailto:haeumkim@illinois.edu}{haeumkim@illinois.edu}
}

\date{}

\maketitle

\begin{abstract}
Computing non-equilibrium steady states of open quantum systems is a challenging task on conventional computers, motivating quantum algorithms for direct steady-state estimation.
A natural route is to regard the steady state as the zero mode of the Liouvillian and to isolate this sector spectrally.
We formulate this task as a known-zero-sector projection problem and implement the corresponding filter using the Rodeo algorithm, which performs stochastic spectral filtering through repeated controlled evolutions and measurement-conditioned filtering steps.
In the steady-state setting, the filter can be centered directly at the known zero eigenvalue, avoiding the spectral search required in generic eigenstate preparation.
Compared with a phase-estimation-based implementation of the same projection, the Rodeo approach enables restart on failure and reduces the target-error dependence of the filtering cost and controlled-evolution depth from power-law to logarithmic.
This advantage becomes more pronounced as the spectral separation $g$ of the Hermitian Liouvillian embedding increases, allowing Rodeo filtering to outperform phase-estimation filtering already at modest controlled-evolution depths.
Our results identify Rodeo filtering as a resource-efficient primitive for estimating steady-state observables in open quantum systems.
\end{abstract}


\section{Introduction}
\label{intro}
Computing non-equilibrium steady states (NESSs) of open quantum systems is a central problem in quantum many-body physics.
In Markovian open-system dynamics, the density matrix evolves under a Liouvillian generator, and a steady state is a density matrix left invariant by this dissipative evolution.
Equivalently, the NESS is characterized as the zero mode of the Liouvillian.
As the object governing the long-time physics of Markovian open systems~\cite{lindblad1976generators,gorini1976completely}, the steady state is central to questions of stability and uniqueness in dissipative dynamics~\cite{baumgartner2008analysis,nigro2019uniqueness}, driven-dissipative phases and dissipative state engineering~\cite{diehl2008quantum,verstraete2009quantum}, quantum transport, and dissipative phase transitions~\cite{sieberer2016keldysh}.
At the same time, computing the steady state on conventional computers is challenging: for an $N$-qubit system, the density matrix lives in a Liouville space whose dimension grows exponentially with $N$, and convergence by direct time evolution can become slow when the Liouvillian gap is small.
These difficulties have motivated several quantum algorithmic approaches to open-system steady states, including equilibrium-state and correlation-function estimation~\cite{terhal2000problem}, variational NESS algorithms~\cite{yoshioka2020variational}, and direct steady-state estimation~\cite{ramusat2021quantum}.

A natural way to formulate direct steady-state estimation on a quantum computer is to isolate the zero sector of a Hermitian Liouvillian embedding.
Following Ref.~\cite{ramusat2021quantum}, the density matrix is vectorized and the generally non-Hermitian Liouvillian is embedded into a Hermitian operator acting on an extended register.
For a unique steady state, this Hermitian embedding has a distinguished zero sector containing both the identity reference mode and the steady-state mode, while all other modes are separated from this sector by an embedding spectral separation, which we denote by $g$.
The algorithmic task is therefore to implement a filter that preserves this zero sector and suppresses the nonzero spectrum.
Phase estimation realizes this projection by resolving the zero eigenvalue and postselecting on the corresponding outcome~\cite{ramusat2021quantum}, in line with broader phase-estimation-based filtering and eigenstate-preparation methods with well-characterized precision guarantees~\cite{wang2023quantum,ding2024quantum,lee2025filtered}.
The drawback is that reducing leakage into the zero bin requires increasing the phase resolution, and hence the maximum controlled-evolution depth, with a power-law dependence on the target filtering error.

The Rodeo algorithm provides a different way to implement such a spectral filter~\cite{choi2021rodeo}.
Rather than estimating an eigenvalue into a phase register, Rodeo acts directly on amplitudes through repeated ancilla-assisted controlled evolutions and success-conditioned measurements.
Eigencomponents at the target eigenvalue are left unchanged, while off-target components acquire oscillatory filtering factors and are progressively suppressed.
Thus Rodeo should be viewed not as a high-resolution eigenvalue readout, but as a measurement-conditioned projection primitive centered at a chosen spectral value.
Recent demonstrations have also established Rodeo filtering as an implementable state-preparation primitive on quantum devices~\cite{qian2024demonstration}.

In this work, we use this observation to formulate a filtering principle for known-zero-sector projection: once the embedding separation from zero is resolved, repeated Rodeo filtering suppresses the residual nonzero-sector weight with only logarithmic dependence on the target filtering error. We apply this principle to the Hermitian Liouvillian embedding by choosing the objective operator to be the embedding itself and centering the filter at the known steady-state eigenvalue, $E=0$. This directly realizes the Liouvillian zero-sector filter without the spectral search required in generic eigenstate preparation. Under the same Liouvillian embedding, input construction, controlled-evolution access, and ratio readout as the phase-estimation approach, this replacement changes the target-error dependence of the filtering primitive from power-law to logarithmic. Restart on failure gives an additional operational advantage, since unsuccessful attempts can be aborted after any filtering step. When the embedding spectral separation is sufficiently large, the resulting separation is visible already at modest controlled-evolution depths. Together, these results identify Rodeo filtering as a resource-efficient primitive for steady-state observable estimation.

\begin{table}[!t]
\centering
\caption{Complexity comparison for the zero-sector filtering step in steady-state estimation.}
\label{tab:complexity_comparison}
\small
\begin{tabularx}{\textwidth}{@{}>{\raggedright\arraybackslash}X
>{\centering\arraybackslash}p{0.34\textwidth}
>{\centering\arraybackslash}p{0.28\textwidth}@{}}
\hline\hline
Quantity 
& Phase-estimation filtering~\cite{ramusat2021quantum} 
& Rodeo filtering \\
\hline
Controlled-evolution depth
&
\(O\!\left(1/(g\varepsilon^{1/2})\right)\)
&
\(O\!\left(\log(1/\varepsilon)/g\right)\)
\\[6pt]

Dominant precision parameter
&
\(O\!\left(\log(1/(g\varepsilon^{1/2}))\right)\)
&
\(O\!\left(\log(1/\varepsilon)\right)\)
\\[6pt]

Gate cost
&
\(O\!\left((2N+1)^k/(g\varepsilon^{1/2})\right)\)
&
\(O\!\left((2N+1)^k\log(1/\varepsilon)/g\right)\)
\\
\hline\hline
\end{tabularx}
\end{table}

\section{Main results}
\label{sec:main_results}

Our first result is an analytic resource separation for the zero-sector filtering primitive used in Liouvillian steady-state estimation.
The comparison is made under the same Liouvillian encoding.
Both methods use the Hermitian Liouvillian embedding $M$, target the known zero eigenvalue, and assume access to controlled time evolution under $M$.
The only change is the mechanism used to isolate the zero sector.
Phase-estimation filtering suppresses leakage by increasing the phase resolution.
Rodeo filtering instead fixes the resolution scale set by the spectral separation and suppresses the remaining nonzero modes through repeated successful filtering steps.

Table~\ref{tab:complexity_comparison} summarizes the analytic comparison. Here $g$ denotes the spectral separation between the zero sector and the remaining spectrum of $M$, and $\varepsilon$ denotes the target filtering error. Constant prefactors are suppressed.

The improvement is localized to the filtering primitive.
At fixed spectral separation $g$, the target-error dependence changes from $O(\varepsilon^{-1/2})$ for phase-estimation filtering to $O(\log(1/\varepsilon))$ for Rodeo filtering.
This improvement does not come from changing the Liouvillian embedding, the input construction, or the observable readout.
It comes from replacing phase-resolution-based projection by measurement-conditioned Rodeo filtering.
Rodeo filtering also provides an operational advantage through restart on failure.
Because the success condition is checked after each filtering step, an unsuccessful attempt can be terminated before the remaining controlled evolutions are executed.
The derivation and assumptions behind Table~\ref{tab:complexity_comparison} are given in Sec.~\ref{sec:resource_comparison}.

\begin{figure}[!t]
\centering
\includegraphics[width=\columnwidth]{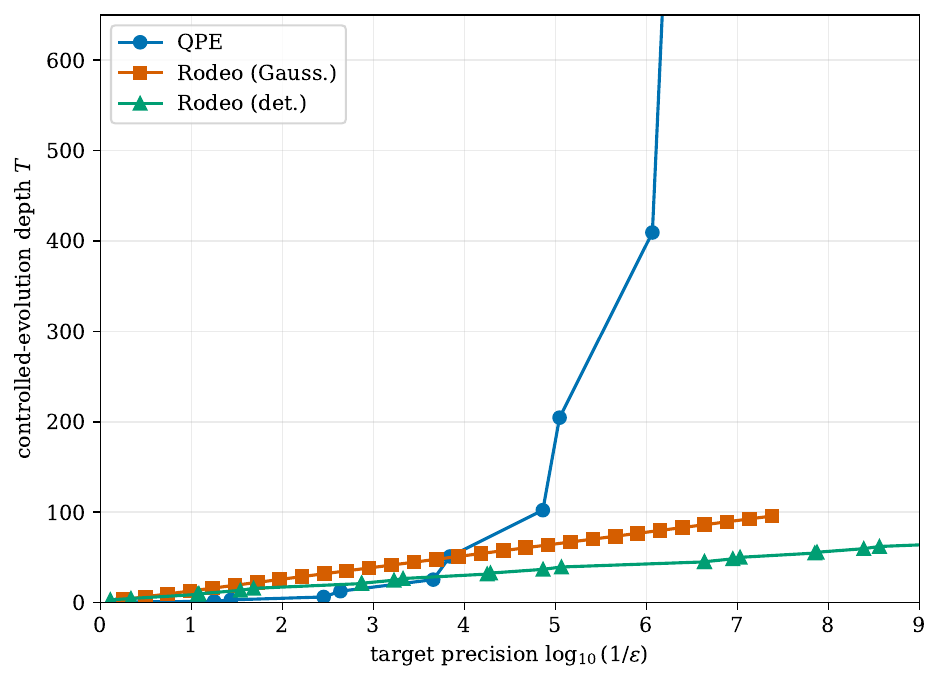}
\caption{Resource cost of the zero-sector filter as a function of the target precision $\log_{10}(1/\varepsilon)$ for the single-spin model ($h=0.5$, $g=1/2$). Since the horizontal axis counts target digits, the approximately linear growth of the Rodeo curves reflects the logarithmic dependence $T_{\rm Rodeo}=O(\log(1/\varepsilon)/g)$, whereas the phase-estimation depth grows rapidly according to its power-law dependence on $1/\varepsilon$. The deterministic low-discrepancy schedule gives a single exact value for each filtering-step count; for the Gaussian schedule, where $\beta_j$ fluctuates from run to run, the precision $\varepsilon$ at each filtering-step count is obtained from the analytic expected residual weight $\mathbb{E}[|\beta_j|^2]$. Gate counts differ only by the common $(2N{+}1)^k$ prefactor.}
\label{fig:cost}
\end{figure}

We support this analytic separation with benchmark simulations on open-system models. Figure~\ref{fig:cost} shows the representative single-spin driven--dissipative benchmark of Ref.~\cite{ramusat2021quantum}, using the same Hermitian embedding and ratio readout for both filters. Since the horizontal axis is $\log_{10}(1/\varepsilon)$, the linear growth of the Rodeo depth directly represents the logarithmic target-error dependence, while the phase-estimation depth grows rapidly with the number of target digits. This numerical behavior mirrors the analytic separation in Table~\ref{tab:complexity_comparison}.

The remaining benchmark results test two further features of the comparison.
First, including restart on failure increases the expected-runtime separation, since phase estimation detects failure only at the final measurement while Rodeo can abort cycle by cycle.
Second, in dissipative transverse-field Ising chains, the cost ratio is controlled primarily by the embedding spectral separation $g$.
The Rodeo advantage grows as the steady-state zero sector becomes better separated from the nonzero spectrum, and it can already appear at modest controlled-evolution depths.
These benchmark results are discussed in detail in Sec.~\ref{sec:benchmark_simulation}.

\section{Preliminaries}

This section collects the quantum-algorithmic primitives needed for our construction. We first recall how steady-state estimation can be formulated as the isolation of the zero-eigenvalue sector of a Hermitian Liouvillian embedding, and then review the Rodeo algorithm as a measurement-based spectral filtering procedure implemented through controlled real-time evolutions.

\subsection{Liouvillian zero-sector filtering for steady-state estimation}
\label{sec:zero_sector_filtering}

Following Ref.~\cite{ramusat2021quantum}, we review a quantum algorithmic framework for direct steady-state estimation based on the zero-eigenvalue sector of a Hermitian Liouvillian embedding.
Let $\mathcal H$ be the finite-dimensional Hilbert space of the system, and let $\mathcal L$ be the Liouvillian generator of the Markovian dynamics,
\begin{align}
    \frac{d\rho}{dt}=\mathcal L(\rho).
\end{align}
A non-equilibrium steady state is a density matrix $\rho_{\rm ss}$ satisfying $\mathcal L(\rho_{\rm ss})=0$. 
Following the operator-vector correspondence in Liouville space, we use the vectorization convention
\begin{align}
    |X\rangle
    :=
    \sum_{jk}X_{jk}|j\rangle\otimes |k\rangle .
\end{align}
The Liouvillian superoperator $\mathcal L$ is then represented by a matrix $L$ on the doubled Hilbert space, so that the steady state is equivalently the right zero mode
\begin{align}
    L|\rho_{\rm ss}\rangle=0.
\end{align}

Since $L$ is generally non-Hermitian, it is embedded into the Hermitian operator
\begin{align}
\label{eq:emb_ham}
    M=
    \begin{pmatrix}
        0 & L\\
        L^\dagger & 0
    \end{pmatrix},
\end{align}
which acts on one auxiliary qubit and the doubled Liouville-space register.
Trace preservation of the density matrix implies that the vectorized identity is a left zero mode of the Liouvillian matrix, namely $L^\dagger \ket{I}=0$. 
Independently, the steady-state condition gives the right-zero-mode equation $L|\rho_{\rm ss}\rangle=0$, where $\rho_{\rm ss}$ denotes the normalized steady-state density matrix.
Therefore $M$ has two distinguished zero modes,
\begin{align}
\label{eq:zero_mode}
    |\eta_0\rangle=|0\rangle|I\rangle,
    \qquad
    |\eta_1\rangle=|1\rangle|\rho_{\rm ss}\rangle .
\end{align}
Throughout this work, we assume that the steady state is unique and that these two vectors exhaust the zero sector of $M$. Thus,
\begin{align}
    \ker M=
    \operatorname{span}\left\{
        |\eta_0\rangle,\,
        |\eta_1\rangle
    \right\}.
\end{align}

The relevant resolution scale for zero-sector filtering is the spectral separation between the zero sector and the remaining eigenmodes of $M$. We denote it by
\begin{align}
    g:=\min_{j\neq 0,1}|\varphi_j|>0,
\end{align}
where $M|\eta_j\rangle=\varphi_j|\eta_j\rangle$ and
$|\eta_0\rangle$ and $|\eta_1\rangle$ span the zero sector as defined in
Eq.~\eqref{eq:zero_mode}.
Thus, $g$ should be understood as the spectral gap of the Hermitian
Liouvillian embedding, or the embedding separation, rather than as a bare
eigenvalue gap of the non-Hermitian Liouvillian itself. Equivalently, it is the
smallest nonzero singular value of the Liouvillian matrix $L$, and it can differ
from the asymptotic Liouvillian decay rate defined from the real parts of the
Liouvillian eigenvalues. This embedding separation is the resolution scale that
must be resolved in order to isolate the steady-state sector.

The remaining algorithmic task is to isolate this zero sector. 
We describe this step abstractly by a successful-branch filter $\mathcal F_0$, which preserves the zero sector up to a common amplitude and suppresses all nonzero modes:
\begin{align}
    \mathcal F_0|\eta_i\rangle
    =
    \lambda_0|\eta_i\rangle,
    \qquad i=0,1,
\end{align}
and
\begin{align}
    \mathcal F_0|\eta_j\rangle
    =
    r_j|\eta_j\rangle,
    \qquad j\neq 0,1 .
\label{eq:filter_epsilon}
\end{align}
Here $\lambda_0$ is the successful-branch amplitude of the zero sector, and $r_j$ is the residual amplitude of the $j$-th nonzero mode. 
Since only the relative suppression $|r_j/\lambda_0|$ matters, we rescale the successful branch and set $\lambda_0=1$. 
The phase-estimation-based projection of Ref.~\cite{ramusat2021quantum} can be viewed as one realization of this abstract filter. 
In the following sections, we introduce Rodeo filtering as an alternative realization and compare the resource requirements of the two implementations.

Given the successful-branch filter $\mathcal F_0$, we construct the input by combining a fixed identity reference state with a normalized trial state.
Let $|\chi\rangle$ be a normalized state on the doubled Liouville-space register, and prepare
\begin{align}
\label{eq:initial_state}
    |\xi\rangle
    =
    \frac{|0\rangle|I\rangle+|1\rangle|\chi\rangle}{\sqrt{2}} .
\end{align}
Writing the second branch in the eigenbasis of $M$ as
\begin{align}
    |1\rangle|\chi\rangle
    =
    c_1|\eta_1\rangle
    +
    \sum_{j\neq 0,1}c_j|\eta_j\rangle ,
\end{align}
the input state becomes
\begin{align}
    |\xi\rangle
    =
    \frac{|\eta_0\rangle+c_1|\eta_1\rangle}{\sqrt{2}}
    +
    \frac{1}{\sqrt{2}}
    \sum_{j\neq 0,1}c_j|\eta_j\rangle .
\end{align}
Conditioned on success, the filter gives the unnormalized output
\begin{align}
    |\widetilde{\psi}_{\rm out}\rangle
    &:=
    \mathcal F_0|\xi\rangle
    \nonumber\\
    &=
    \frac{|\eta_0\rangle+c_1|\eta_1\rangle}{\sqrt{2}}
    +
    \frac{1}{\sqrt{2}}
    \sum_{j\neq 0,1}c_j r_j|\eta_j\rangle .
    \label{eq:filtered_output}
\end{align}
We write the normalized output as
\( |\psi_{\rm out}\rangle
=|\widetilde{\psi}_{\rm out}\rangle/\sqrt{p_{\rm succ}} \),
with success probability \(p_{\rm succ}=\langle\widetilde{\psi}_{\rm out}|
\widetilde{\psi}_{\rm out}\rangle\).
This expression separates the desired zero-sector component from the residual nonzero-mode contribution. 
The fixed \(|0\rangle|I\rangle\) component provides a reference zero-mode component, while the \(|1\rangle|\chi\rangle\) component carries the steady-state signal through its overlap \(c_1\) with \(|\eta_1\rangle\).

The remaining step is to extract the steady-state observable from the coherence between the two zero-sector branches. 
For compactness, denote the readout of a system observable $\hat O$ by
\begin{align}
    R_O
    :=
    \big\langle\psi_{\rm out}\big| 
    X_{\rm aux}\otimes \hat O\otimes \Id
    \big|\psi_{\rm out}\big\rangle.
\label{eq:readout_RO}
\end{align}
where \(X_{\rm aux}\) acts on the auxiliary qubit. 
With the vectorization convention above,
\begin{align}
    \langle I|(\hat O\otimes \Id)|\rho_{\rm ss}\rangle
    =
    \operatorname{Tr}(\hat O\rho_{\rm ss}).
\label{eq:vec_trace_identity}
\end{align}
To quantify the size of the residual nonzero-sector contamination, we define
\begin{align}
    r_{\max}:=\max_{j\neq0,1}|r_j|.
\end{align}
Here $r_{\max}$ is the largest residual amplitude assigned by the successful-branch filter to any nonzero eigenmode, after normalizing the zero-sector amplitude to one.
It therefore controls the error induced by imperfect zero-sector filtering.
Combining Eqs.~\eqref{eq:readout_RO} and \eqref{eq:vec_trace_identity} with the filtered output in Eq.~\eqref{eq:filtered_output}, the raw readout becomes
\begin{align}
    R_O
    =
    \frac{1}{p_{\rm succ}}
    \operatorname{Re}
    \left[
        c_1\,\operatorname{Tr}(\hat O\rho_{\rm ss})
    \right]
    +
    \mathcal O(r_{\max}) .
\end{align}

The unknown overlap $c_1$ and output normalization $p_{\rm succ}$ are common to the readouts for $\hat O$ and $\hat O=\Id$. 
Taking their ratio, with the same interference quadrature, yields
\begin{align}
    \frac{R_O}{R_I}
    =
    \operatorname{Tr}(\hat O\rho_{\rm ss})
    +
    \mathcal O(r_{\max}),
    \label{eq:observable_ratio}
\end{align}
where $\operatorname{Tr}(\rho_{\rm ss})=1$.

\subsection{Rodeo algorithm}
\label{sec:rodeo_algorithm}

The Rodeo algorithm is a stochastic spectral filtering method for preparing eigenstates of a Hermitian operator near a chosen target eigenvalue.
Let $H_{\rm obj}$ be a Hermitian operator with eigenpairs $\{(E_j,|E_j\rangle)\}$, and let $E$ be the target eigenvalue. 
Starting from an initial state
\begin{align}
    |\psi_0\rangle=\sum_j a_j |E_j\rangle,
\end{align}
the algorithm applies a sequence of measurement-conditioned filtering steps.
In the $\ell$-th step, a time $t_\ell$ is chosen, a controlled evolution under $H_{\rm obj}$ is applied with an ancilla, and the ancilla is measured.
Conditioned on obtaining the designated success outcome, the procedure acts as a spectral filter centered at $E$.

For a single filtering step with evolution time $t$, an eigencomponent $|E_j\rangle$ is multiplied, up to an irrelevant phase convention, by
\begin{align}
    \cos\!\left(\frac{(E_j-E)t}{2}\right).
\end{align}
After $n$ successful filtering steps with times $t_1,\ldots,t_n$, the corresponding filtering factor is
\begin{align}
    \beta_j
    =
    \prod_{\ell=1}^{n}
    \cos\!\left(\frac{(E_j-E)t_\ell}{2}\right).
\label{eq:rodeo_beta_general}
\end{align}
Thus eigenstates with $E_j=E$ are preserved, while components with $E_j\neq E$ are suppressed.

The times $t_\ell$ are sampled independently from a Gaussian distribution with root-mean-square width $t_{\rm rms}$. 
This time scale sets the spectral resolution of the filter: eigenvalues separated from the target by more than order $1/t_{\rm rms}$ acquire rapidly fluctuating phases and are suppressed under repeated successful filtering steps. 
For such off-target components, the typical residual weight decreases geometrically with the number of filtering steps, giving a target-error dependence that is logarithmic in the desired suppression.

A useful feature of the Rodeo algorithm is that the success condition is checked after every filtering step.
If a failure outcome is obtained at an intermediate step, the attempt can be terminated and restarted without executing the remaining steps. 
In the following sections, this filtering mechanism is applied to the Hermitian Liouvillian embedding by setting $H_{\rm obj}=M$ and centering the filter at the known steady-state eigenvalue $E=0$.

\section{Rodeo realization of the Liouvillian zero-sector filter}
\label{sec:rodeo_liouvillian}

As shown in Sec.~\ref{sec:zero_sector_filtering}, steady-state estimation reduces to implementing a successful-branch filter $\mathcal F_0$ that preserves the zero sector of the Hermitian Liouvillian embedding $M$ and suppresses all nonzero modes. We realize this filter using the Rodeo algorithm reviewed in Sec.~\ref{sec:rodeo_algorithm}. The mapping is direct: the Rodeo objective operator is chosen as $H_{\rm obj}=M$, and the filter is centered at the known target eigenvalue $E=0$. Thus, unlike generic Rodeo eigenstate preparation, no scan over candidate target energies is required; the desired sector is fixed by construction as the zero-eigenvalue sector of $M$.

With this specialization, the Rodeo filtering factor becomes explicit. For an eigenmode $M|\eta_j\rangle=\varphi_j|\eta_j\rangle$, $n$ successful filtering steps give
\begin{align}
\beta_j
=
\prod_{\ell=1}^{n}
\cos\left(\frac{\varphi_j t_\ell}{2}\right).
\label{eq:rodeo_beta_liouvillian}
\end{align}
The two zero modes satisfy $\varphi_0=\varphi_1=0$ and are therefore left invariant, $\beta_0=\beta_1=1$. Nonzero modes acquire the residual factors $\beta_j$. Comparing this action with the abstract filter of Sec.~\ref{sec:zero_sector_filtering}, the Rodeo realization is obtained by identifying
\begin{align}
r_j=\beta_j,
\qquad j\neq0,1.
\label{eq:epsilon_beta}
\end{align}
This is the basic replacement that turns the abstract zero-sector filter into an explicit Rodeo filter.

The required number of Rodeo filtering steps follows from the suppression of these residual factors. For the Gaussian-time version used as a baseline, Appendix~\ref{app:rodeo_suppression} gives the averaged bound
\begin{align}
\mathbb{E}\left[|\beta_j|^2\right]
\le
q^n,
\qquad q<1,
\qquad j\neq0,1,
\label{eq:rodeo_suppression_bound}
\end{align}
provided the sampling time scale resolves the spectral separation $g$. Hence obtaining residual weight at most $\varepsilon$ requires only
\begin{align}
n=O\left(\log(1/\varepsilon)\right)
\end{align}
successful Rodeo filtering steps. In this sense, the Rodeo implementation inherits the logarithmic target-error dependence of the standard Rodeo filter.

\begin{algorithm}[t]
\captionof{algorithm}{Rodeo filtering for Liouvillian steady-state estimation}
\label{alg:rodeo_steady_state}
\vspace{0.4em}
\hrule
\vspace{0.6em}
\begin{algorithmic}[1]
\Require Liouvillian matrix \(L\); observable \(\hat O\); target filtering error \(\varepsilon\); Rodeo time schedule \(\{t_\ell\}_{\ell=1}^{n}\) resolving the spectral separation \(g\)
\Ensure Estimate of \(\langle \hat O\rangle_{\rm ss}=\operatorname{Tr}(\hat O\rho_{\rm ss})/\operatorname{Tr}(\rho_{\rm ss})\)

\State Construct the Hermitian Liouvillian embedding \(M\) via Eq.~\eqref{eq:emb_ham}.
\State Prepare the input state \(|\xi\rangle\) via Eq.~\eqref{eq:initial_state}.
\State Set the target eigenvalue \(E\gets 0\).
\Statex \textit{Filtering stage}
\For{\(\ell=1,\ldots,n\)}
    \State Select \(t_\ell\) from the prescribed schedule.
    \State Prepare a Rodeo ancilla in \(|1\rangle\), then apply \(H\).
    \State Apply controlled-\(e^{-iMt_\ell}\) to the \(M\)-register.
    \State Apply the phase rotation \(P(E t_\ell)\) to the Rodeo ancilla.
    \State Apply \(H\) to the Rodeo ancilla and measure it.
    \If{the designated success outcome is not obtained}
        \State Abort and restart from \(|\xi\rangle\).
    \EndIf
\EndFor
\Statex \textit{Readout stage}
\State Estimate \(R_O=\langle X_{\rm aux}\otimes(\hat O\otimes \Id)\rangle\) on the successful output state.
\State Estimate \(R_I\) by repeating the readout with \(\hat O=\Id\).
\State \Return \(R_O/R_I\).
\Statex
\Statex \textit{Notes:} Since \(E=0\), the target phase rotation is trivial. The schedule may be Gaussian random or deterministic geometric.
\end{algorithmic}
\vspace{0.6em}
\hrule
\end{algorithm}

The Gaussian-time implementation, however, has an additional sampling issue: the residual factors $\beta_j$ fluctuate from run to run because the evolution times are randomly drawn. This fluctuation affects the stability of the filtering step, although not the zero-sector mapping itself. Optimized Rodeo variants replace the random Gaussian times by deterministic schedules, such as geometric time schedules, which remove the random-time component of the residual-factor fluctuations and can improve suppression constants~\cite{cohen2023optimizing,patkowski2026improved}. These optimized schedules can be incorporated into the present framework by replacing $\beta_j$ with the corresponding deterministic filtering factors. Thus the depth scaling in the target error remains logarithmic, but the filtering implementation becomes more stable.

A further operational feature inherited from the Rodeo algorithm is restartability.
The filter is implemented as a sequence of measurement-conditioned filtering steps, so a failure outcome can be detected before the full sequence has been executed.
When this occurs, the current attempt can be terminated and restarted from the initial state, avoiding the remaining controlled evolutions in that failed run.
This restart structure does not change the conditioned successful branch, but it contributes to the expected runtime analyzed in Sec.~\ref{sec:resource_comparison}.

Thus the abstract filter of Sec.~\ref{sec:zero_sector_filtering} is made explicit by the replacement $r_j=\beta_j$, or by the corresponding deterministic residual factor for an optimized schedule.
The resulting steady-state estimator is the same ratio readout in Eq.~\eqref{eq:observable_ratio}.
The full procedure is summarized in Algorithm~\ref{alg:rodeo_steady_state}.

\section{Resource comparison with phase-estimation filtering}
\label{sec:resource_comparison}

We now compare the resource cost of implementing the zero-sector filter using phase estimation and Rodeo filtering.
The comparison is restricted to the filtering primitive: both methods use the same Hermitian Liouvillian embedding $M$, target the same zero sector, and assume access to controlled time evolution under $M$.
We write $g$ for the spectral separation between the zero sector and the nonzero spectrum of $M$, and $\varepsilon$ for the target filtering error.

In the phase-estimation approach, the zero sector is isolated by resolving eigenvalues of $M$ and postselecting the phase-estimation outcome corresponding to zero.
The finite phase register produces a filter with nonzero leakage: an eigenmode separated from zero by at least $g$ can still be assigned to the zero bin with a probability controlled by the phase resolution.
For a maximum controlled-evolution time $T_{\rm QPE}$, this leakage scales as $O(1/(g^2T_{\rm QPE}^2))$.
Requiring this leakage to be at most $\varepsilon$ gives
\begin{align}
T_{\rm QPE}
=
O\left(\frac{1}{g\varepsilon^{1/2}}\right).
\end{align}
Equivalently, the required phase-register size, or number of controlled time-evolution powers, scales as
\begin{align}
m_{\rm QPE}
=
O\left(
\log\left(\frac{1}{g\varepsilon^{1/2}}\right)
\right).
\end{align}
The detailed leakage estimate is given in Appendix~\ref{app:qpe_filtering_estimate}.

In the Rodeo approach, the spectral resolution and the suppression error are controlled separately.
The evolution times are chosen on the scale $1/g$ so that the zero sector is resolved from the nonzero spectrum.
Once this resolution scale is reached, additional successful Rodeo filtering steps multiply each nonzero mode by residual factors whose weight decreases exponentially with the number of steps.
Consequently, achieving target filtering error $\varepsilon$ requires only
\begin{align}
n_{\rm Rodeo}
=
O\left(\log(1/\varepsilon)\right)
\end{align}
successful filtering steps, as derived in Appendix~\ref{app:rodeo_suppression}.
The corresponding total controlled-evolution depth is therefore
\begin{align}
T_{\rm Rodeo}
=
O\left(\frac{\log(1/\varepsilon)}{g}\right).
\end{align}

To convert these controlled-evolution depths into gate counts, we use the same simulation model for both methods.
Specifically, we assume a Suzuki--Trotter implementation of the controlled evolution under the $(k+1)$-local Hermitian operator $M$ acting on $2N+1$ qubits.
Under the normalization $\|M\|=O(1)$ and suppressing the simulation-accuracy overhead, the cost of simulating total time $T$ scales as $O((2N+1)^kT)$; see Appendix~\ref{app:gate_cost_model}.
This gives
\begin{align}
G_{\rm QPE}
&=
O\left(
\frac{(2N+1)^k}{g\varepsilon^{1/2}}
\right),
\\
G_{\rm Rodeo}
&=
O\left(
\frac{(2N+1)^k\log(1/\varepsilon)}{g}
\right).
\end{align}

The comparison shows that the improvement is localized to the filtering primitive.
Phase-estimation filtering reduces leakage by increasing phase resolution, which ties the target error directly to the maximum evolution time.
Rodeo filtering instead fixes the resolution scale at $O(1/g)$ and reduces the residual nonzero-mode weight through repeated successful filtering steps.
This changes the target-error dependence of the filtering step from $O(\varepsilon^{-1/2})$ to $O(\log(1/\varepsilon))$.

\section{Benchmark simulations on open-system models}
\label{sec:benchmark_simulation}

We use two complementary benchmarks.
First, we compare Rodeo and phase-estimation filtering on the single-spin
driven--dissipative model used in Ref.~\cite{ramusat2021quantum}.
This provides a direct check of the target-error scaling and of the observable
ratio readout.
Second, we use dissipative transverse-field Ising chains to vary the embedding
spectral separation $g$ and test how the cost ratio depends on this resolution
scale.

\subsection{Single-spin benchmark}
\label{sec:single_spin_benchmark}

We begin with the spin-$1/2$ driven--dissipative model of
Ref.~\cite{ramusat2021quantum}. The system is subject to a transverse field
$\hat H=h\,\hat\sigma_x$ and relaxation along $z$, with single jump operator
$\hat A=\hat\sigma^-$, so that
\begin{align}
\dot{\rho}
= -i\big[\hat H,\rho\big]
-\tfrac{1}{2}\big\{\hat\sigma^+\hat\sigma^-,\rho\big\}
+\hat\sigma^-\rho\,\hat\sigma^+ .
\end{align}
We vectorize the density matrix, construct the Liouvillian matrix $L$, and form
the Hermitian embedding $M$ of Eq.~\eqref{eq:emb_ham}. For this model the
embedding has the expected two-dimensional zero sector of
Eq.~\eqref{eq:zero_mode}, and its spectral separation is $g=1/2$, independent
of the field strength $h$.

\begin{figure}[tbp]
\centering
\includegraphics[width=\columnwidth]{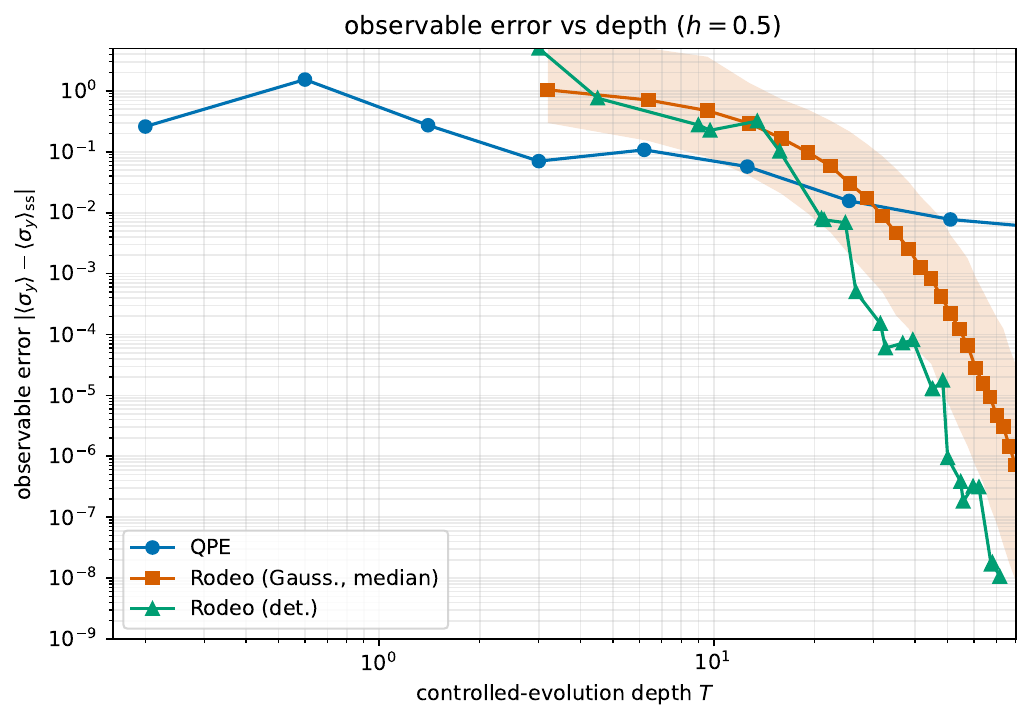}
\caption{Single-shot observable error of $\langle\hat\sigma_y\rangle$ versus
controlled-evolution depth at $h=0.5$ (log--log). The phase-estimation filter
is shown as a line; for the Gaussian schedule we plot the median single-shot
error with its $10$--$90\%$ spread over random schedules. The deterministic
schedule is exact. Both Rodeo schedules cross below the phase-estimation curve
at moderate depth and continue to fall, while phase estimation plateaus,
reflecting the logarithmic versus power-law filtering cost.}
\label{fig:obs_error}
\end{figure}

Both filters act on the same embedding $M$, the same input state $|\xi\rangle$,
and the same ratio readout of Eq.~\eqref{eq:observable_ratio}. The only
difference is the filtering primitive. We quantify the filtering error
$\varepsilon$ as the worst-case residual weight of a nonzero mode, as in
Table~\ref{tab:complexity_comparison}: for phase estimation this is computed
from the zero-bin leakage amplitudes, while for Rodeo it is computed from the
residual factors of Eq.~\eqref{eq:rodeo_beta_liouvillian}. Thus
$\varepsilon$ is a filtering error, not the observable error itself.

The controlled-evolution depth is the total simulated time under $M$. For phase
estimation this is $T_{\rm QPE}=t_0(2^m-1)$, with $t_0=1/5$ as in
Ref.~\cite{ramusat2021quantum}. For Rodeo it is
$T_{\rm Rodeo}=\sum_\ell |t_\ell|$. We compare a Gaussian schedule with
$\trms=\kappa/g$ and $\kappa=2$ against a deterministic low-discrepancy schedule
whose times are bounded at the scale $1/g$. Throughout this benchmark we use
the real, phase-symmetrized successful-branch filters, so that residual phases
from a literal controlled-$e^{-iMt}$ implementation do not affect the
residual-weight cost comparison. The corresponding compiled dynamic circuit and
a Trotterized realization of the single-spin controlled evolution are described
in the Supplemental Material.

The benchmark summarized in Fig.~\ref{fig:cost} confirms the analytic scaling
of Table~\ref{tab:complexity_comparison}. A fit to the data gives
$T_{\rm QPE}\propto\varepsilon^{-0.51}$, consistent with the predicted
$O(\varepsilon^{-1/2})$ dependence, while both Rodeo schedules scale linearly in
$\log(1/\varepsilon)$. The fitted exponents are stable under changes of the
constant time-scale parameters; these constants affect the prefactors and
crossover location, but not the functional form of the scaling. At
$\varepsilon=10^{-8}$, for example, the deterministic Rodeo schedule reduces the
controlled-evolution depth from $6553.4$ for phase estimation to $59.6$, a
factor of about $110$.

Figure~\ref{fig:obs_error} verifies that the same resource separation is visible
at the level of physical observables. For the single-spin model, both Rodeo
schedules reduce the single-shot observable error below the phase-estimation
curve at moderate controlled-evolution depth, while the phase-estimation error
plateaus at the corresponding finite-resolution value. The full convergence of
$\langle\hat\sigma_y\rangle$ and $\langle\hat\sigma_z\rangle$ for
$h\in\{0.5,1.0,1.5\}$ is shown in Fig.~\ref{fig:observables} of
Appendix~\ref{app:observable_convergence}.

\subsection{Expected runtime with restart on failure}
\label{sec:restart}

The depths above count a single successful filtering run.
We now include failed attempts and restarts.
For the normalized input used here, the total zero-sector weight is
\(4/7\simeq0.571\), so both filters have the same resolved-regime success
probability \(p_{\rm run}\to4/7\).
The difference is the cost of failure.
Phase estimation detects failure only at the terminal measurement, whereas
Rodeo can abort after the first failed filtering step.

For the deterministic Rodeo schedule, we model the restart process exactly.
The expected total depth to obtain one successful filtered output is
\begin{align}
\mathbb{E}[T_{\rm Rodeo}^{\rm tot}]
= \frac{1}{p_{\rm run}}\sum_{r=1}^{n}\big(W_{r-1}-W_r\big)\,D_r + D_n,
\end{align}
where
\begin{align}
W_r
=
\sum_j w_j
\prod_{s\le r}
\cos^2\!\left(\frac{\varphi_j t_s}{2}\right),
\qquad
w_j=|c_j|^2,
\end{align}
and $D_r=\sum_{\ell\le r}|t_\ell|$ is the depth through cycle $r$.
For phase estimation, a failed attempt costs the full circuit depth, giving the
expected depth $T_{\rm QPE}/p_{\rm run}$.

The resulting restart overhead for the deterministic Rodeo schedule remains
small over the range considered here, never exceeding $\sim1.1$ and decreasing
toward unity as more cycles are used. By contrast, phase estimation pays the
fixed overhead $1/p_{\rm run}$. Combined with the single-run depth scaling of
Fig.~\ref{fig:cost}, this increases the expected-total-depth advantage at
$\varepsilon=10^{-8}$ from about $110$ to roughly $1.8\times10^2$. A
circuit-level illustration of the same early-abort mechanism is provided in the
Supplemental Material. This benchmark assumes ideal postselection statistics
and a fault-tolerant setting; finite-shot readout noise and device errors are
outside its scope.

\subsection{Dependence on the spectral separation}
\label{sec:separation_dependence}

The single-spin benchmark fixes the spectral separation at $g=1/2$. To test the
dependence on $g$ directly, we use the dissipative transverse-field Ising chain
of Ref.~\cite{ramusat2021quantum},
\begin{align}
\hat H
&=
\frac{J}{4}\sum_{\langle j,k\rangle}\hat Z_j\hat Z_k
+
\frac{h}{2}\sum_j \hat X_j,
\\
\hat A_j&=\sqrt{\gamma}\,\hat\sigma_j^- .
\end{align}
We sweep the dissipation strength $\gamma$, interaction strength $J$, and system
size $N$. For each instance, we construct the Hermitian embedding $M$, compute
its spectral separation $g=\min_{j\neq0,1}|\varphi_j|$, and compare the
gate-cost ratio $G_{\rm QPE}/G_{\rm Rodeo}$ at fixed controlled-evolution depth
using the same residual-weight metric as above.

\begin{figure}[tbp]
\centering
\includegraphics[width=\columnwidth]{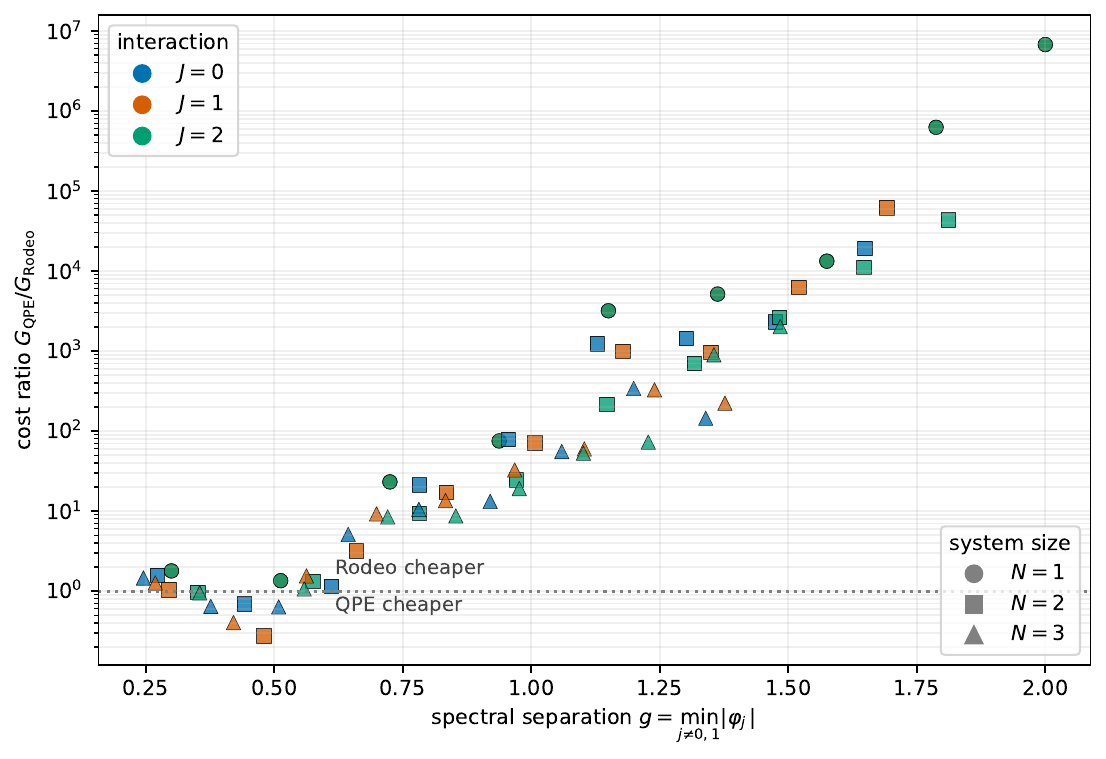}
\caption{Gate-cost ratio $G_{\rm QPE}/G_{\rm Rodeo}$ at fixed controlled-evolution depth, as a function of the spectral separation $g$ of the embedding $M$. Each point is a dissipative transverse-field Ising chain, swept over dissipation strength, interaction strength $J$ (color), and system size $N$ (marker). The data are organized primarily by $g$, with the Rodeo advantage increasing with $g$. The dotted line marks equal cost. The single-spin benchmark corresponds to $g=1/2$.}
\label{fig:cost_vs_g}
\end{figure}

Figure~\ref{fig:cost_vs_g} shows that the cost ratio is organized primarily by
the embedding separation $g$. The ratio $G_{\rm QPE}/G_{\rm Rodeo}$ increases
with $g$, crosses unity near $g\approx0.5$ for the fixed depth used here, and
reaches several orders of magnitude for $g\gtrsim1$. The same $g$-organized
trend persists in interacting instances, as shown in
Appendix~\ref{app:sep_interacting}. Thus the Rodeo advantage at modest
controlled-evolution depth is largest when the zero sector is well separated
from the nonzero spectrum of the Hermitian embedding.

This dependence should be interpreted in terms of the embedding separation,
rather than the asymptotic Liouvillian decay rate. Appendix~\ref{app:sep_vs_decay} shows that the embedding separation $g$, not the
asymptotic decay rate $g_{\rm decay}$, is the variable that organizes the cost ratio. Appendix~\ref{app:sep_density} further shows that
spectral density gives only a finite-spectrum correction at fixed $g$. These
diagnostics support the interpretation that larger embedding separation is the
dominant factor behind the observed Rodeo advantage.

\section{Conclusion}
\label{sec:conclusion}

We have introduced a Rodeo-based realization of the Liouvillian zero-sector filter for direct steady-state estimation in open quantum systems.
The construction uses the same Hermitian Liouvillian embedding, input state, and ratio readout as the phase-estimation approach of Ref.~\cite{ramusat2021quantum}, but replaces the phase-estimation projection by measurement-conditioned Rodeo filtering.
Because the steady-state eigenvalue is known to be zero, the Rodeo filter can be centered directly at the target sector without a spectral search.
At the level of the filtering primitive, this replacement changes the target-error dependence of the controlled-evolution depth and gate cost from power-law to logarithmic.
It also provides an operational advantage through restart on failure, since unsuccessful attempts can be aborted before the remaining filtering steps are executed.
Our benchmark simulations confirm these features on open-system models and show that the advantage becomes more pronounced when the zero sector is well separated from the nonzero spectrum of the Hermitian embedding.

The main conceptual lesson is that steady-state estimation is not only a Liouvillian-encoding problem, but also a spectral-filtering problem.
Whenever the desired eigenvalue is fixed by structure, phase-estimation-based projection is not the only natural filtering primitive.
The Liouvillian steady state provides a particularly clean example because the target eigenvalue is exactly zero by definition.
This suggests that Rodeo filtering may be useful more broadly in quantum algorithms that use phase estimation mainly as a projection tool rather than as a high-resolution eigenvalue readout.
Examples include structured zero-mode, nullspace, or symmetry-sector projection problems where the target spectral value is known in advance.
In such settings, replacing phase-estimation filtering by Rodeo filtering can reduce the precision dependence while retaining the same encoded problem and readout strategy.

At the circuit level, Rodeo filtering is also well matched to recent progress in dynamic-circuit hardware.
Phase estimation can already be made more qubit-efficient when the final inverse quantum Fourier transform is replaced by semiclassical or dynamic-circuit implementations using measurement and feed-forward~\cite{griffiths1996semiclassical,corcoles2021exploiting,baumer2024quantum}.
Iterative variants can even reduce phase estimation to a single ancillary qubit~\cite{dobsicek2007arbitrary}.
These developments show that mid-circuit measurement can be a useful resource for spectral algorithms, not only for error correction.
Recent work on measuring, mitigating, and preserving quantum information during mid-circuit measurement and reset further clarifies the hardware requirements for using such primitives reliably~\cite{hothem2025measuring,motlakunta2024preserving}.
Rodeo filtering uses this capability in a more direct way.
The measurement ancilla is not used to reconstruct phase bits or to implement a Fourier transform; it is itself the filtering element.
This makes Rodeo especially natural for projection tasks with a known target eigenvalue, such as the Liouvillian zero mode studied here.

To turn this filtering-level advantage into a practical steady-state algorithm, the next step is to make the full implementation pipeline model adapted.
The first requirement is an efficient realization of the Hermitian Liouvillian embedding and its controlled time evolution.
For local Lindbladians and dissipative circuits, this points toward embeddings, dilation-based constructions, and simulation methods that exploit locality, sparsity, or channel structure rather than treating the embedded generator as a generic Hermitian operator.
The second requirement is to optimize the Rodeo layer under realistic circuit constraints, including deterministic time schedules, adaptive restart policies, finite-shot ratio readout, and imperfect mid-circuit measurements.
Finally, practical applications will require understanding regimes beyond the unique, well-separated steady state assumed here, such as degenerate steady-state manifolds, near-degenerate zero sectors, and small-gap dissipative dynamics.
These directions would determine when the filtering-level resource separation established in this work becomes an end-to-end advantage for steady-state estimation on quantum devices.

\section{Acknowledgement}

For reproducibility, all numerical simulations presented in this work have been reimplemented and made available at Ref.~\cite{kim_rodeo_ness_code_2026}.
This work was supported by the education and training program of the Quantum Information Research Support Center, funded through the National research foundation of Korea (NRF) by the Ministry of science and ICT (MSIT) of the Korean government(No. RS-2023-NR057243), and the National Research Council of Science \& Technology (NST) (Grant No. GTL25011-401). H.Y. acknowledges support from the National Research Foundation of Korea (NRF) through a grant funded by the Ministry of Science and ICT (Grant No. RS-2025-00515537), the Institute for Information \& Communications Technology Promotion (IITP) grants funded by the Korean government (MSIP) (Grants Nos. RS2019-II190003 and RS-2025-02304540), and the Korea Institute of Science and Technology Information (Grant No. P26028). J.J. acknowledges support from the National Research Foundation of Korea (NRF) through a grant funded by the Ministry of Science and ICT (Grant No. RS-2023-00211817).

\bibliographystyle{unsrt}
\bibliography{reference}

\clearpage
\appendix

\section{Filtering and gate-cost estimates}
\label{app:filtering_gate_cost}
This appendix collects the estimates underlying the resource comparison in
Table~\ref{tab:complexity_comparison}. We first derive the residual-weight
suppression for Rodeo and phase-estimation filtering, and then convert the
corresponding controlled-evolution depths into gate-count scalings using a
common simulation model.

\subsection{Rodeo filtering estimate}
\label{app:rodeo_suppression}

We derive the Rodeo entries in Table~\ref{tab:complexity_comparison}. 
We use the Gaussian-time version of the Rodeo filter as the baseline implementation and specialize it to the Hermitian Liouvillian embedding by setting $H_{\rm obj}=M$ and $E=0$. 
The off-target energy separation in the standard Rodeo estimate is then replaced by the spectral separation $g$ between the zero sector of $M$ and its nonzero spectrum.

For a nonzero mode $M|\eta_j\rangle=\varphi_j|\eta_j\rangle$, the residual amplitude after $n$ successful Rodeo filtering steps is
\begin{align}
\beta_j
=
\prod_{\ell=1}^{n}
\cos\left(\frac{\varphi_j t_\ell}{2}\right),
\qquad j\neq0,1,
\end{align}
where $t_\ell\sim\mathcal N(0,t_{\rm rms}^2)$.
Averaging the residual weight for a single filtering step gives
\begin{align}
\mathbb E_t
\left[
\cos^2\left(\frac{\varphi_j t}{2}\right)
\right]
=
\frac{1+
e^{-\varphi_j^2t_{\rm rms}^2/2}}
{2}.
\end{align}
By independence of the sampled times,
\begin{align}
\mathbb E\left[|\beta_j|^2\right]
=
\left[
\frac{1+
e^{-\varphi_j^2t_{\rm rms}^2/2}}
{2}
\right]^n .
\label{eq:app_rodeo_suppression}
\end{align}

Since $|\varphi_j|\ge g$ for all nonzero modes, choosing $t_{\rm rms}=\kappa/g$ with fixed $\kappa>0$ gives
\begin{align}
\mathbb E\left[|\beta_j|^2\right]
\le
q(\kappa)^n,
\qquad
q(\kappa)=\frac{1+e^{-\kappa^2/2}}{2}<1 .
\end{align}
Thus the averaged residual weight is exponentially suppressed in the number of successful Rodeo filtering steps.
To make this residual weight at most $\varepsilon$, it suffices to take
\begin{align}
n
=
O\left(\log(1/\varepsilon)\right).
\end{align}

The expected magnitude of a Gaussian evolution time is
\begin{align}
\mathbb E[|t_\ell|]
=
t_{\rm rms}\sqrt{\frac{2}{\pi}}
=
O(1/g).
\end{align}
Therefore the total controlled-evolution depth of the Rodeo filter scales as
\begin{align}
T_{\rm Rodeo}
=
O\left(\frac{\log(1/\varepsilon)}{g}\right).
\end{align}

\subsection{Phase-estimation filtering estimate}
\label{app:qpe_filtering_estimate}

We derive the phase-estimation entries in Table~\ref{tab:complexity_comparison}. 
The estimate follows the standard phase-estimation implementation of the zero-sector projection.
Let $M|\eta_j\rangle=\varphi_j|\eta_j\rangle$, with $|\eta_0\rangle$ and $|\eta_1\rangle$ spanning the zero sector, and assume that every nonzero mode satisfies $|\varphi_j|\ge g$.

Let $m$ be the number of phase-register qubits. 
For an eigenmode whose dimensionless phase is $\phi_j\neq0$, the amplitude assigned to the zero outcome is
\begin{align}
\alpha_0^{(j)}
=
\frac{1}{2^m}
\frac{1-e^{2\pi i2^m\phi_j}}
{1-e^{2\pi i\phi_j}}.
\end{align}
Thus
\begin{align}
|\alpha_0^{(j)}|^2
=
\frac{1}{2^{2m}}
\frac{\sin^2(\pi2^m\phi_j)}
{\sin^2(\pi\phi_j)}.
\end{align}
For nonzero modes separated from the zero sector by $g$, this leakage is bounded, up to constant prefactors, by
\begin{align}
|\alpha_0^{(j)}|^2
=
O\left(\frac{1}{g^2 2^{2m}}\right).
\end{align}
Requiring this leakage to be at most $\varepsilon$ gives
\begin{align}
2^m
=
O\left(\frac{1}{g\varepsilon^{1/2}}\right),
\end{align}
or equivalently
\begin{align}
m
=
O\left[
\log\left(\frac{1}{g\varepsilon^{1/2}}\right)
\right].
\end{align}
The largest controlled-evolution time in phase estimation scales as $T_{\rm QPE}=O(2^m)$, so
\begin{align}
T_{\rm QPE}
=
O\left(\frac{1}{g\varepsilon^{1/2}}\right).
\end{align}

\subsection{Gate-cost conversion}
\label{app:gate_cost_model}

We estimate the gate cost using a Suzuki--Trotter simulation of the controlled evolution $e^{-iMt}$.
For a $(k+1)$-local Hermitian operator $M$ acting on $2N+1$ qubits, the $p$-th order Suzuki--Trotter expansion approximates $e^{-iMT}$ with simulation error $\delta_{\rm sim}$ using
\begin{align}
N_{\rm gates}
=
O\left(
\frac{
(2N+1)^k
\|M\|^{1+1/p}
T^{1+1/p}
}
{\delta_{\rm sim}^{1/p}}
\right)
\end{align}
elementary gates~\cite{childs2021theory}.
Assuming $\|M\|=O(1)$ and taking $p$ sufficiently large, this reduces to the scaling
\begin{align}
N_{\rm gates}
=
O\left((2N+1)^k T\right),
\end{align}
up to the chosen simulation-accuracy overhead.

Substituting the phase-estimation depth gives
\begin{align}
G_{\rm QPE}
&=
O\left(
(2N+1)^k T_{\rm QPE}
\right)
\\
&=
O\left(
\frac{(2N+1)^k}{g\varepsilon^{1/2}}
\right).
\end{align}
Substituting the Rodeo depth gives
\begin{align}
G_{\rm Rodeo}
&=
O\left(
(2N+1)^k T_{\rm Rodeo}
\right)
\\
&=
O\left(
\frac{(2N+1)^k\log(1/\varepsilon)}{g}
\right).
\end{align}

\section{Observable convergence for all field values}
\label{app:observable_convergence}

\begin{figure}[h]
\centering
\includegraphics[width=0.92\textwidth]{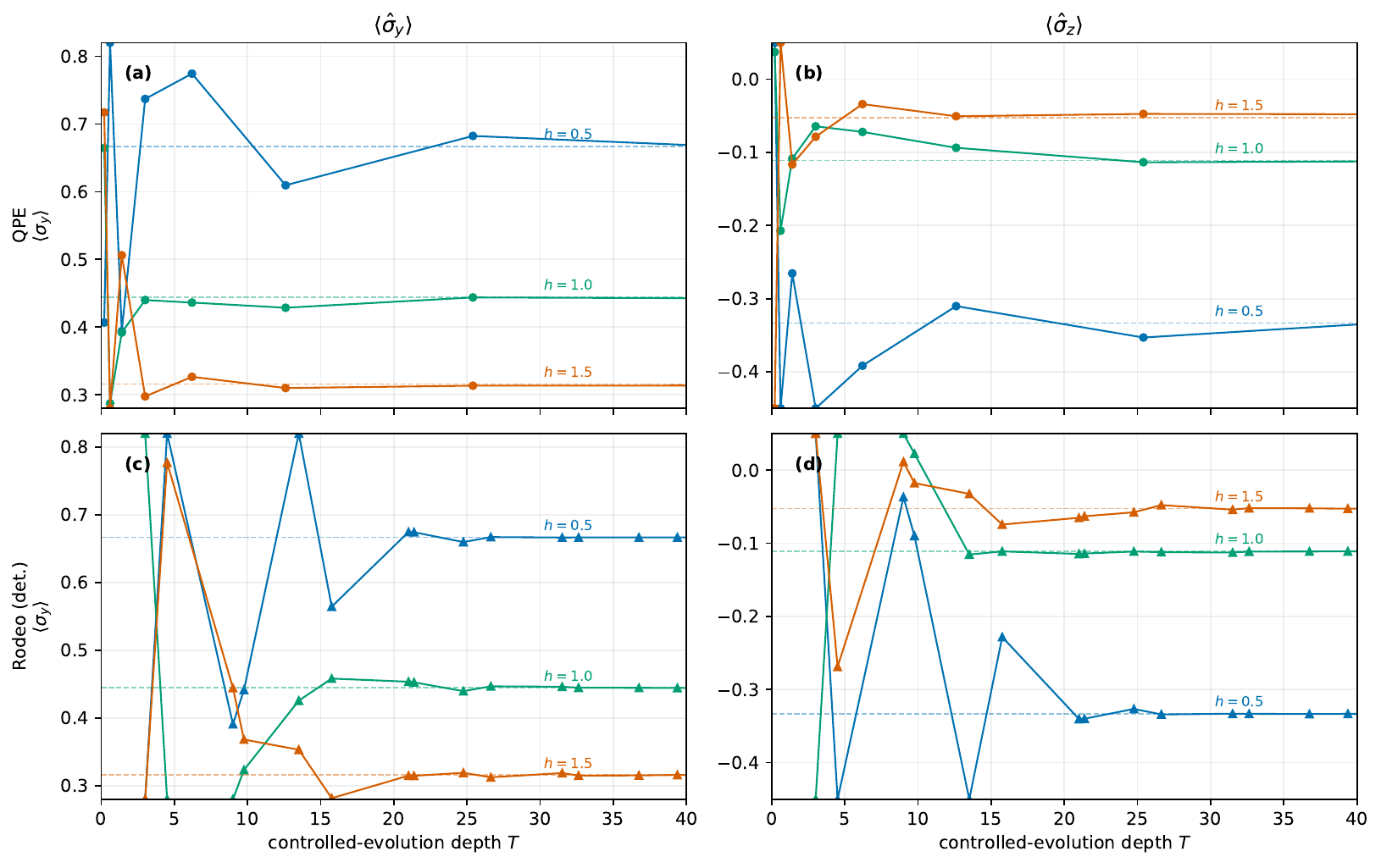}
\caption{Steady-state observables across the field values $h\in\{0.5,1.0,1.5\}$, with
each $h$ drawn in a distinct color. Top row [(a),(b)]: phase estimation (circles);
bottom row [(c),(d)]: deterministic Rodeo filtering (triangles). Left column [(a),(c)]:
$\langle\hat\sigma_y\rangle$; right column [(b),(d)]: $\langle\hat\sigma_z\rangle$.
These are the two physically independent non-trivial observables of the model, since
$\langle\hat\sigma_x\rangle=0$ identically. The estimates obtained from the ratio
readout converge to the distinct exact NESS values (dashed lines) as the
controlled-evolution depth $T$ increases. The vertical axes are restricted to the
convergence window; at small depth the ratio estimator swings strongly because the
normalization readout passes near zero.}
\label{fig:observables}
\end{figure}

For completeness, Fig.~\ref{fig:observables} shows the convergence of both non-trivial steady-state observables, $\langle\hat\sigma_y\rangle$ and $\langle\hat\sigma_z\rangle$, for all three field values $h\in\{0.5,1.0,1.5\}$ and for both the phase-estimation and the deterministic-Rodeo filters. This is the accuracy check summarized in the main text by Fig.~\ref{fig:obs_error}.

\section{Separation dependence of the filter cost}
\label{app:separation}

This appendix collects the numerical evidence behind the claim of
Sec.~\ref{sec:separation_dependence} that the spectral separation $g$ governs the
Rodeo-versus-phase-estimation cost comparison. All quantities are computed by exact
diagonalization of the Hermitian embedding $M$; the filter costs are the worst-case
residual weights $\max_{j\neq0,1}|\beta_j|^2$ (Rodeo) and
$\max_{j\neq0,1}|\alpha_0^{(j)}|^2$ (phase estimation) at fixed controlled-evolution
depth, with the Rodeo schedule the deterministic low-discrepancy sequence of
Sec.~\ref{sec:benchmark_simulation} and the phase register the best admissible
register at the same depth. The model is the dissipative transverse-field Ising
chain of Ref.~\cite{ramusat2021quantum}, swept over dissipation strength $\gamma$,
interaction strength $J$, and system size $N$.

\subsection{Embedding separation versus Liouvillian decay rate}
\label{app:sep_vs_decay}

The embedding separation $g=\min_{j\neq0,1}|\varphi_j|$ is the smallest nonzero
singular value of the Liouvillian matrix $L$. This quantity differs in general
from the asymptotic Liouvillian decay rate
$g_{\rm decay}=\min_{\lambda\neq0}|\re\,\lambda|$, defined from the nonzero
Liouvillian eigenvalues. The two coincide for the single-spin model
($g=g_{\rm decay}=1/2$), but they need not coincide for interacting or
non-normal Liouvillian dynamics.

\begin{figure}[h]
\centering
\includegraphics[width=\columnwidth]{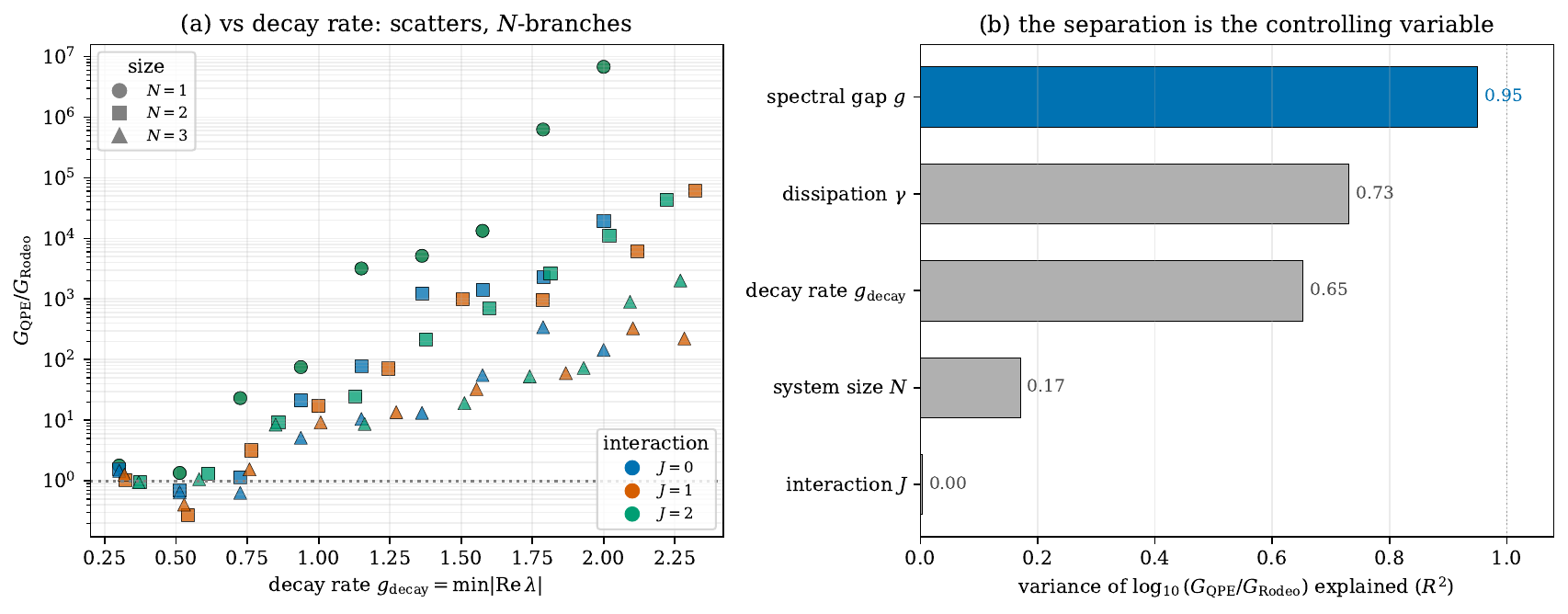}
\caption{Cost ratio $G_{\rm QPE}/G_{\rm Rodeo}$ at fixed controlled-evolution depth
over a $(\gamma,N,J)$ grid. (a) When plotted against the Liouvillian decay rate
$g_{\rm decay}$, the data scatter and separate systematically with system size.
(b) Fraction of the variance of $\log_{10}(G_{\rm QPE}/G_{\rm Rodeo})$ explained by each
control variable alone (single-variable $R^2$): the spectral separation $g$ explains
$0.95$, far more than the dissipation $\gamma$ ($0.73$), the decay rate $g_{\rm decay}$
($0.65$), the system size $N$ ($0.17$), or the interaction $J$ ($<0.01$). This
identifies $g$, rather than the asymptotic decay rate, as the relevant resolution scale
for the filter-cost comparison.}
\label{fig:app_collapse}
\end{figure}

Figure~\ref{fig:app_collapse} quantifies which scale controls the cost ratio. Panel
(a) plots the ratio against the Liouvillian decay rate $g_{\rm decay}$: the data
scatter and separate by system size. Panel (b) reports, for each candidate variable,
the fraction of the variance of $\log_{10}(G_{\rm QPE}/G_{\rm Rodeo})$ that it explains
on its own; the embedding separation $g$ accounts for far more ($0.95$) than the decay
rate $g_{\rm decay}$ ($0.65$) or any other parameter. This supports the use of $g$,
rather than the asymptotic decay rate, as the resolution scale controlling the
Rodeo-versus-phase-estimation filtering comparison.

\subsection{Organization by $g$ survives interactions}
\label{app:sep_interacting}
\begin{figure}[h]
\centering
\includegraphics[width=\columnwidth]{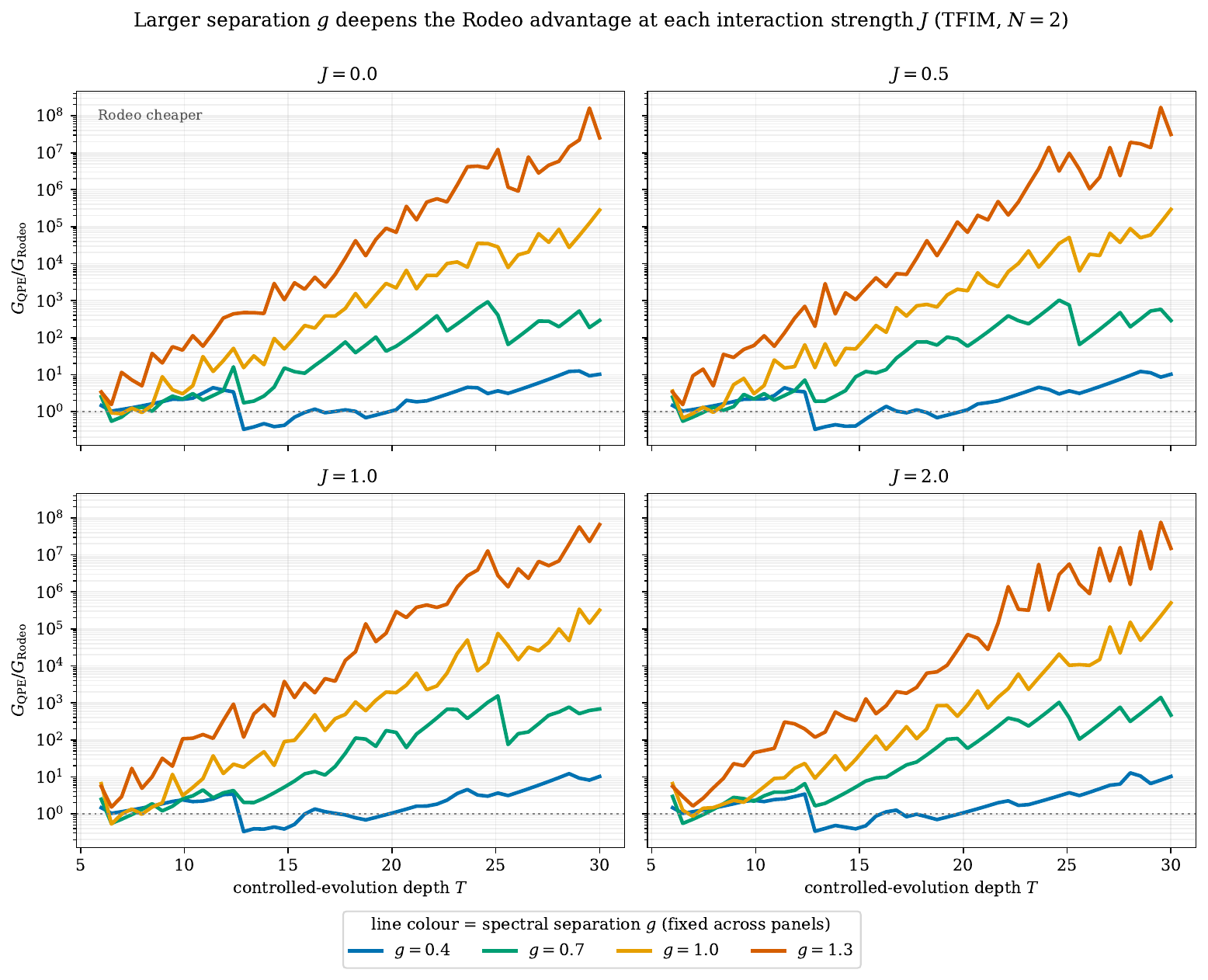}
\caption{The Rodeo advantage is organized by the spectral separation $g$ at every
interaction strength (dissipative transverse-field Ising model, $N=2$). Each panel
fixes the interaction $J\in\{0,0.5,1,2\}$; within a panel the cost ratio
$G_{\rm QPE}/G_{\rm Rodeo}$ is plotted versus controlled-evolution depth $T$ for four
fixed separations $g\in\{0.4,0.7,1.0,1.3\}$ (one color each, the same four values in
every panel). The same-color curve barely moves from panel to panel: at fixed $g$ the
advantage and its growth with $T$ are essentially the same from $J=0$ to $J=2$, so the
interaction is not an independent control once $g$ is fixed.}
\label{fig:app_interacting}
\end{figure}

The organization by $g$ is not an artifact of the non-interacting limit.
Figure~\ref{fig:app_interacting} repeats the experiment with the Ising coupling
switched on, displayed as a $2\times2$ grid over the interaction strength
$J\in\{0,0.5,1,2\}$. Within every panel the cost ratio is drawn for the same four
fixed separations $g$ (one color each); the curve at a given $g$ barely shifts from
panel to panel, so at fixed $g$ the Rodeo advantage and its growth with depth $T$ are
essentially the same across interaction strengths. Neither the interaction strength
nor the system size is an independent control once $g$ is fixed.

\subsection{Spectral density as a finite-spectrum correction}
\label{app:sep_density}

The filtering factors act on the full embedding spectrum $\{\varphi_j\}$ rather
than only on the closest nonzero mode. Thus, at fixed minimum separation $g$,
one may ask whether the number of nonzero modes near the zero sector provides an
additional control parameter for the cost comparison. To isolate this effect
from changes in $g$, we use synthetic spectra in which the smallest separation
is fixed while the number of embedding modes above it is varied.

\begin{figure}[h]
\centering
\includegraphics[width=\columnwidth]{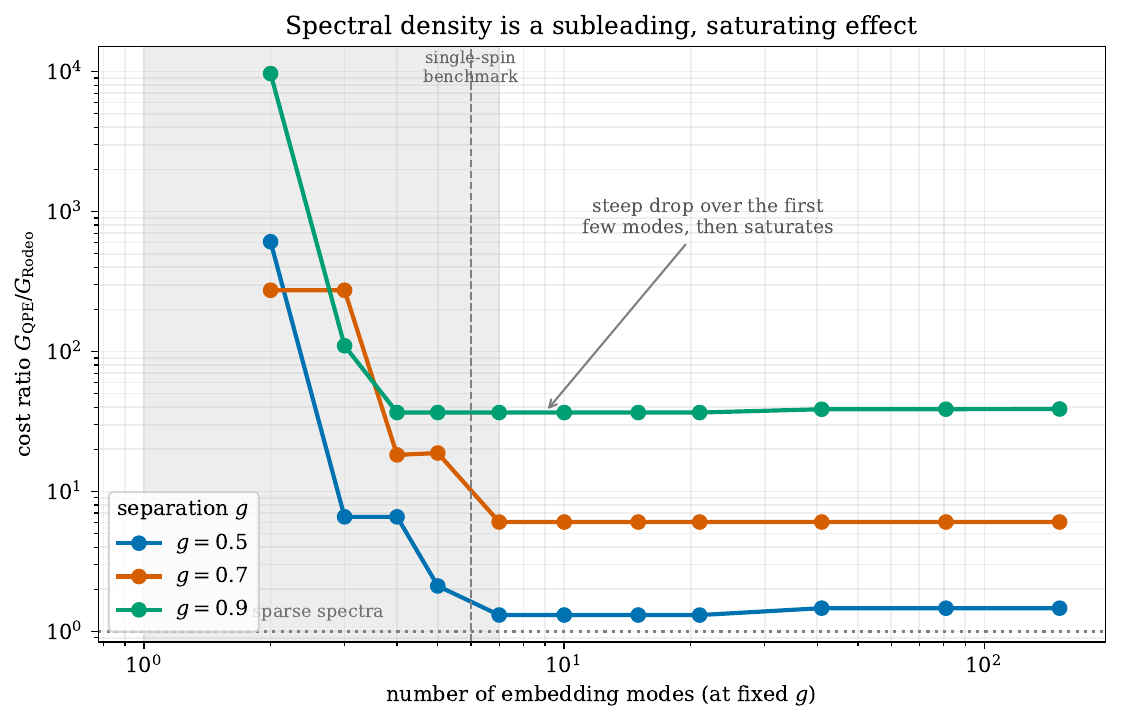}
\caption{Spectral-density dependence at fixed embedding separation, using
synthetic spectra. Increasing the number of embedding modes changes the cost
ratio most strongly for the first few modes and then saturates. The separation
$g$ sets the overall scale of the Rodeo advantage, while the spectral density
contributes a finite-spectrum correction.}
\label{fig:app_density}
\end{figure}

Figure~\ref{fig:app_density} shows that this spectral-density dependence is
visible but limited. At fixed $g$, increasing the number of embedding modes
reduces the Rodeo advantage over the first few modes and then saturates, with
little further change once several modes are present. By contrast, changing $g$
moves the cost ratio over orders of magnitude. Thus the embedding separation
remains the main organizing scale of the comparison: larger $g$ gives a larger
Rodeo advantage at fixed controlled-evolution depth, while spectral density
enters only as a finite-spectrum correction.

\newpage

\renewcommand{\thesection}{S\arabic{section}}
\renewcommand{\thefigure}{S\arabic{figure}}
\renewcommand{\theequation}{S\arabic{equation}}

\begin{center}
{\Large\bf Supplemental Material:}
{\Large Rodeo Filtering for Direct Steady-State Estimation in Open Quantum Systems}
\end{center}

\section{Overview}
\label{supp:overview}

The main manuscript analyzes the Liouvillian zero-sector filter at the level of successful-branch residual amplitudes.
Both phase-estimation filtering and Rodeo filtering act on the same Hermitian Liouvillian embedding, and the resource comparison is made before committing to a particular low-level compilation of the controlled time evolution.
This supplement collects the corresponding small-scale circuit checks for the single-spin benchmark.

The checks have three roles.
First, they show how the phase-symmetric Rodeo filter used in the benchmark can be represented as a measurement-conditioned dynamic circuit.
Second, they connect the dense single-spin controlled unitary to a Pauli-string Trotter decomposition of the same embedding.
Third, they confirm on a noiseless statevector simulator that the post-selected ratio readout reproduces the expected steady-state observable and that failed Rodeo attempts can abort before the remaining controlled evolutions are executed.
Device noise, imperfect mid-circuit measurement, finite-fidelity reset, and scalable block-encoding optimizations are outside the scope of this supplement.

\section{Single-spin embedding and phase-symmetric Rodeo step}
\label{supp:setup}

For the single-spin benchmark, the Liouvillian matrix $L$ is embedded into the Hermitian operator
\begin{align}
    M=
    \begin{pmatrix}
        0 & L\\
        L^\dagger & 0
    \end{pmatrix},
    \label{eq:supp_embedding}
\end{align}
which acts on one branch qubit and the doubled Liouville-space register.
Since the doubled Liouville space is four dimensional for a single spin, the embedding acts on a three-qubit data register.
The two distinguished zero modes are the identity-reference mode and the steady-state mode,
\begin{align}
    \ket{\eta_0}=\ket{0}\ket{I},
    \qquad
    \ket{\eta_1}=\ket{1}\ket{\rho_{\rm ss}}.
    \label{eq:supp_zero_modes}
\end{align}
The input state used in the benchmark is
\begin{align}
    \ket{\xi}
    =
    \frac{\ket{0}\ket{I}+\ket{1}\ket{\chi}}{\sqrt{2}},
    \label{eq:supp_input}
\end{align}
where $\ket{\chi}$ is a normalized trial state on the doubled Liouville-space register.

A successful Rodeo step based directly on controlled-$e^{-iMt}$ multiplies an eigenmode of $M$ by a real filtering factor together with an eigenvalue-dependent phase.
The phase does not affect residual weights, but it can affect finite-depth observable readouts if left uncompensated.
For the benchmark filters and for the compiled circuit below, we therefore use the phase-symmetric branch implementation.
The two ancilla branches apply $e^{+iMt/2}$ and $e^{-iMt/2}$, so that the success outcome implements the real operator
\begin{align}
    \cos\!\left(\frac{Mt}{2}\right).
    \label{eq:supp_cos_filter}
\end{align}
For an eigenmode $M\ket{\eta_j}=\varphi_j\ket{\eta_j}$, $n$ successful cycles produce the residual factor
\begin{align}
    \beta_j=
    \prod_{\ell=1}^{n}
    \cos\!\left(\frac{\varphi_j t_\ell}{2}\right),
    \label{eq:supp_beta}
\end{align}
with $\beta_0=\beta_1=1$ for the zero sector.
Thus the circuit implements the same real successful-branch filter used in the residual-weight analysis of the main text.

\section{Dynamic Rodeo circuit}
\label{supp:dynamic_circuit}

We place the Hermitian embedding $M$ on a three-qubit data register
$\mathsf{m}=(\text{row},\text{col},\text{branch})$
with one reused Rodeo ancilla $\mathsf{r}$.
For this small model, the controlled evolutions are implemented as dense controlled unitaries, which is exact but specific to the single-spin benchmark.
A scalable implementation would replace these dense unitaries by a Hamiltonian-simulation or block-encoding construction adapted to the structure of $M$.

\begin{figure}[tbp]
\centering
\includegraphics[width=0.92\textwidth]{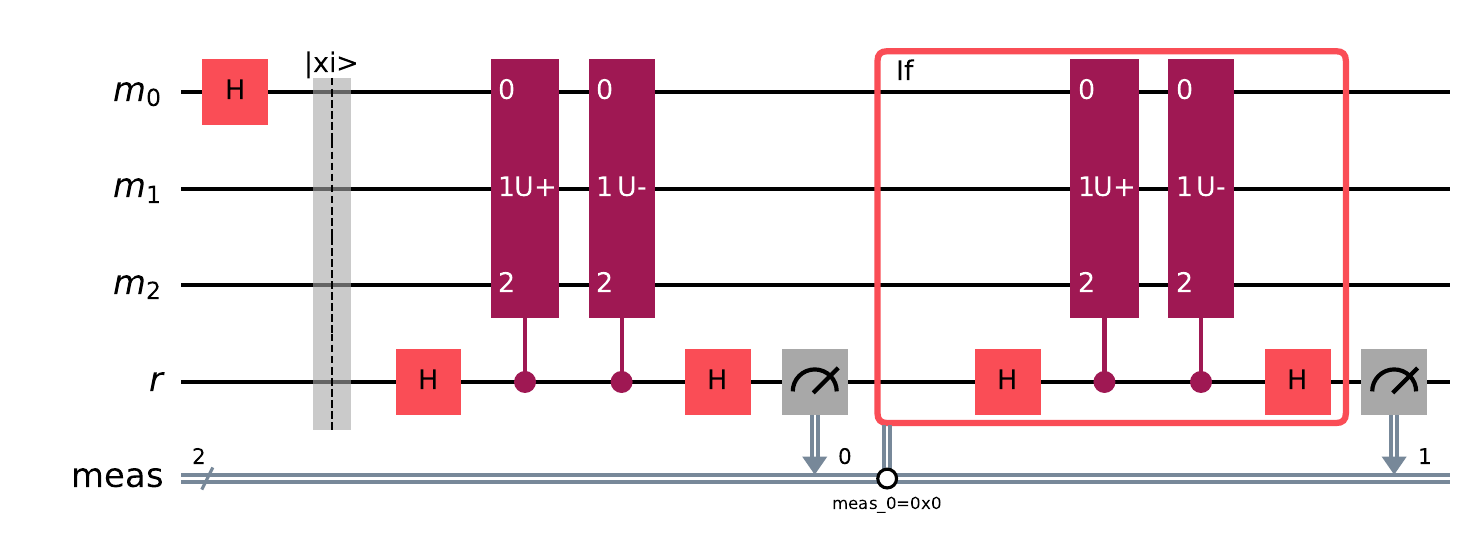}
\caption{Dynamic Rodeo filter circuit for the single-spin embedding $M$, shown for two
cycles ($n=2$) and drawn from the compiled Qiskit circuit. The data register is
$\mathsf{m}=(\text{row},\text{col},\text{branch})$, the Rodeo ancilla is reused as
$\mathsf{r}$, and the success bits are stored in the classical register
\texttt{meas}. The controlled-evolution gates are
$U_{\pm}\equiv e^{\pm iM t_\ell/2}$. Each cycle applies $H$ on the ancilla, the
symmetric evolutions $U_{\pm}$, a second $H$, and a mid-circuit measurement. Later
cycles sit inside a classically controlled block, so a failed cycle aborts the
remaining controlled evolutions. Each success-bit measurement is placed outside the
conditional, as required by current dynamic-circuit restrictions, and failed shots are
removed by post-selection.}
\label{fig:supp_aer_circuit}
\end{figure}

The defining feature of the compiled circuit is that the filter is built from measurement-conditioned cycles.
After the ancilla of cycle $k$ is measured, the controlled evolutions of later cycles are placed inside a classically controlled block that runs only if all previous success bits have been obtained.
Therefore a failure short-circuits the remaining schedule and the later controlled evolutions are not executed in that shot.
Figure~\ref{fig:supp_aer_circuit} shows two such cycles, with the second cycle nested in the classical condition that tests the first success bit.
We use the deterministic low-discrepancy schedule from the main benchmark for the times $\{t_\ell\}$, the input state $\ket{\xi}$ defined above, and the same ratio readout for observables.
All simulations use $4\times10^{4}$ noiseless shots, so the only stochastic deviation from the ideal filter is shot noise.

\section{Trotterized realization of the controlled evolution}
\label{supp:trotter}

\begin{figure}[tbp]
\centering
\includegraphics[width=0.92\textwidth]{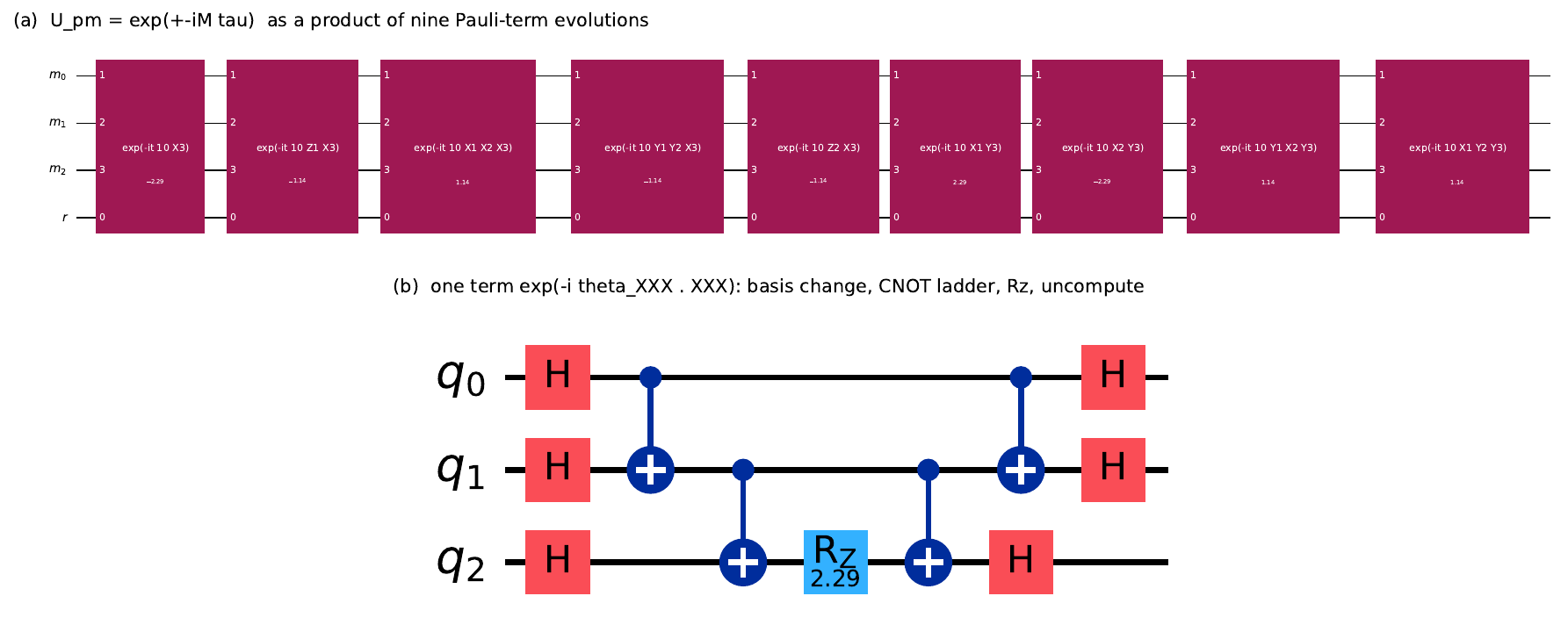}
\caption{Trotter realization of one Rodeo controlled evolution for the single-spin
embedding $M$, drawn from the compiled Qiskit circuit. The controlled evolution
$U_{\pm}=e^{\pm iM\tau}$ is written as a product of Pauli-term evolutions
$e^{-i\theta_P P}$ with
$P\in\{XII,XIZ,XXX,XYY,XZI,YIX,YXI,YXY,YYX\}$. Each term expands into basis changes,
a CNOT ladder, an $R_z$ phase rotation, and the uncomputing ladder.}
\label{fig:supp_aer_trotter}
\end{figure}

The dense unitary $U_{\pm}=e^{\pm iM\tau}$ used in the statevector circuit is exact for the single-spin benchmark, but it is not a scalable primitive.
To connect the circuit picture to the gate-cost model used in the main text, Fig.~\ref{fig:supp_aer_trotter} shows the same controlled evolution as a Trotter compilation.
In the Pauli basis, the single-spin embedding can be written as
\begin{align}
    M=\sum_P c_P P,
    \label{eq:supp_pauli_expansion}
\end{align}
with real coefficients $c_P$.
For this model, the relevant Pauli strings are
\begin{align}
P\in\{XII,XIZ,XXX,XYY,XZI,YIX,YXI,YXY,YYX\}.
\label{eq:supp_pauli_strings}
\end{align}
A first-order Trotter step writes one controlled evolution as an ordered product of single-term evolutions $e^{-i\theta_P P}$ with $\theta_P=c_P\tau$.
Expanding one such term gives the standard pattern of basis changes, a CNOT ladder, one $R_z$ rotation carrying the evolution angle, and the uncomputing ladder.
This is the local Pauli-string primitive underlying the gate-cost conversion in the main text.

\section{Convergence to the exact steady state}
\label{supp:convergence}

\begin{figure}[tbp]
\centering
\includegraphics[width=0.98\textwidth]{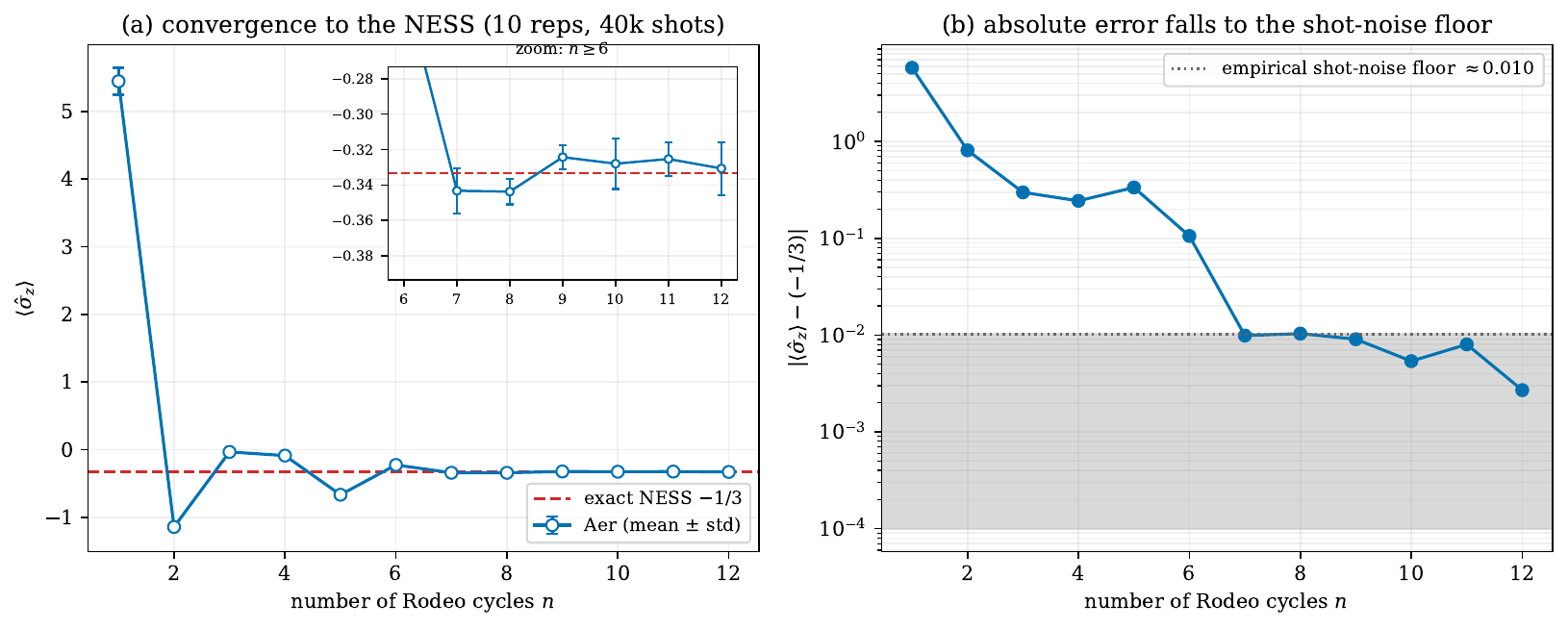}
\caption{Statevector circuit check on the single-spin model ($h=0.5$, $g=1/2$).
(a) The post-selected ratio estimate of $\langle\hat\sigma_z\rangle$ from the compiled
dynamic Rodeo circuit converges to the exact NESS value as the number of cycles $n$
increases. The transient oscillation at small $n$ reflects incomplete,
sign-alternating suppression of nonzero modes by the finite schedule. (b) The absolute
error on a logarithmic scale falls to the shot-noise floor, approximately $10^{-2}$ for
$4\times10^{4}$ shots, once the spectral separation is resolved.}
\label{fig:supp_aer_convergence}
\end{figure}

Figure~\ref{fig:supp_aer_convergence} shows the post-selected estimate of $\langle\hat\sigma_z\rangle$ as a function of the number of Rodeo cycles $n$.
The estimate approaches the exact steady-state value $\langle\hat\sigma_z\rangle_{\rm ss}=-1/3$, and the absolute error falls to the shot-noise floor once the schedule resolves the spectral separation.
We verified that the statevector values agree with a direct evaluation of the product filter $\prod_\ell\cos(Mt_\ell/2)$ on the input state to within the same shot-noise floor.
The approach is non-monotonic at small $n$, because a finite partial product of cosine factors suppresses the nonzero modes incompletely and the residual amplitudes can alternate in sign as cycles are added.
This finite-schedule oscillation is reproduced by the noiseless circuit and decays as additional cycles accumulate.
The same readout with $\hat O=\hat\sigma_y$ converges to $\langle\hat\sigma_y\rangle_{\rm ss}=2/3$.

\section{Gate-level early abort}
\label{supp:early_abort}

\begin{figure}[tbp]
\centering
\includegraphics[width=0.98\textwidth]{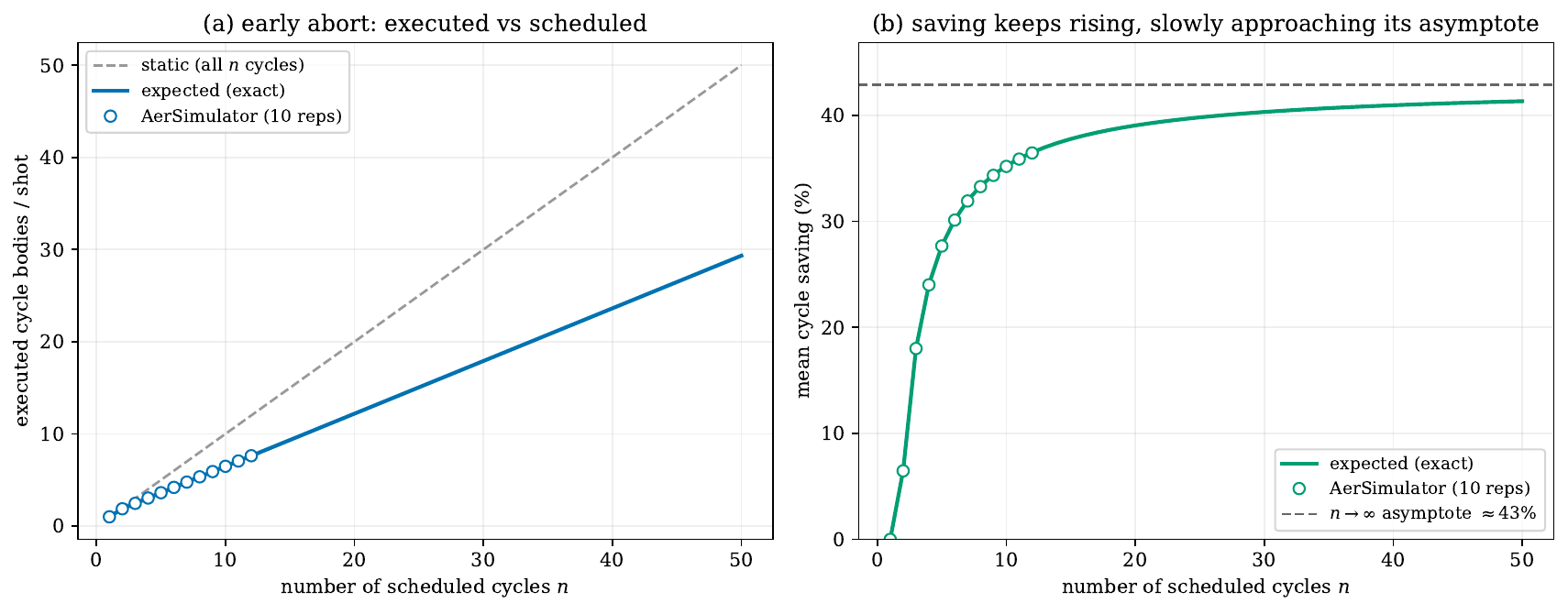}
\caption{Gate-level early-abort saving for the dynamic Rodeo circuit on a noiseless
statevector simulator ($h=0.5$). The mean cycle saving grows with the scheduled depth
and saturates near $25\%$ for this model and schedule. The effect follows from the
measurement-conditioned circuit structure and illustrates the circuit-level mechanism
behind restart on failure.}
\label{fig:supp_aer_cycle_saving}
\end{figure}

The dynamic circuit also exhibits the restart mechanism directly.
For each shot, we record the number of Rodeo cycle bodies executed before the first failed success bit.
The mean of this quantity gives the average executed depth, which we compare in Fig.~\ref{fig:supp_aer_cycle_saving} with the static cost of running all $n$ cycles unconditionally.
The average executed depth grows more slowly than the scheduled depth, so the mean cycle saving from early abort increases with $n$ and saturates near $25\%$ for this model and schedule.
This is the circuit-level counterpart of the restart-on-failure advantage analyzed in the main text: failures are detected cycle by cycle and abort before the controlled evolutions of the remaining cycles are executed.
The saving is structural, following from the measurement-conditioned circuit and single-run success statistics; it is independent of device noise in this noiseless simulation.

Together, Figs.~\ref{fig:supp_aer_convergence} and~\ref{fig:supp_aer_cycle_saving} show that, for the single-spin benchmark, the compiled phase-symmetric Rodeo circuit reproduces the expected steady-state readout in the resolved regime and realizes the early-abort mechanism at the gate level.
These checks support the consistency of the circuit construction, while the asymptotic resource comparison remains the filtering-level analysis presented in the main text.

\end{document}